\documentclass[aps,twocolumn,superscriptaddress,showpacs]{revtex4}
\usepackage{graphicx}
\usepackage{amsmath}
\usepackage{amsfonts}

\begin{document} 

\title{Search for bottleneck effects in Penna ageing and Schulze language model}
\author{Krzysztof Malarz}
\homepage{http://home.agh.edu.pl/malarz/}
\email{malarz@agh.edu.pl}
\affiliation{Faculty of Physics and Applied Computer Science,
AGH University of Science and Technology,\\
al. Mickiewicza 30, PL-30059 Krak\'ow, Euroland}
\author{Dietrich Stauffer}
\email{stauffer@thp.uni-koeln.de} 
\affiliation{Faculty of Physics and Applied Computer Science,
AGH University of Science and Technology,\\
al. Mickiewicza 30, PL-30059 Krak\'ow, Euroland}
\affiliation{visiting from Institute of Theoretical Physics,
University of Cologne, D-50923 K\"oln, Euroland}

\begin{abstract}
No influence was seen when in two models with memory
effects the populations were drastically decreased after equilibrium was 
established, and then allowed to increase again.
\end{abstract}

\maketitle

\section{Introduction}

In reversible models like the Ising paramagnet above the Curie temperature,
the equilibrium distribution is independent of the initial distribution.
This is not always the case in systems with memory, for example if a small
fraction of elements always remains at the initial states. In biology, at 
a ``bottleneck'' (see e.g. \cite{radomski}) most of the population dies out
e.g. due to an environmental catastrophe, and then the population grows back
to about its old size. Immediately after this bottleneck, the distribution of 
genes then is an equilibrium distribution, which in general differs
from both a random distribution and the case where all genomes are identical.
Then, if the system has long-time memory, the final genetic distribution
long after the bottleneck can be different from the one immediately before
the bottleneck.

The present note searches unsuccessfully for such bottleneck effects in the
Penna model for biological ageing and the Schulze model for human languages.
Though both models are reviewed e.g. in \cite{book} we define them in Section
\ref{sec-model} and present our simulations in Section \ref{sec-simulation},
with conclusions in Section \ref{sec-summary}.

\section{Models}
\label{sec-model}

In the Penna model \cite{book,penna}, individuals age according to a bit-string
representing their genome. A bit set to zero represents health, a bit set to 
one represents an inherited disease which starts at the age corresponding to
the position of this 1 within the 32 bits of the string. Three or more 
active diseases kill the individual. At and above the reproductive age of eight
(in time units of bit positions) each individual has three offspring having
the same genome except that one randomly selected bit is set to one. If it
is already one it stays at one. This latter assumption makes the model 
irreversible: Bits set to one never can become zero again. Initially, all bits 
are set to zero.
 
The Schulze model also started with bit-strings, but later \cite{schulze} it 
was generalized to $F$ features each of which can be 1, 2, 3, $\dots$, $Q$, with
$F=8, \; Q=5$ in our simulations below. This string of small integers represents
the language (or grammar) of the individual. Each site of a square lattice
carries one individual, which on each sweep through the lattice is replaced 
by a child having the same language except for one mutation. And also at each
such iteration each individual switches to the language of a randomly 
selected neighbour with probability $0.9 (1-x)^2$ where $x$ is the fraction
of people in the whole population speaking the individual's language. The 
mutations happen for each of the $F$ features with probability $p$; if 
a mutation happens, then with probability 1/2 the feature gets a random 
value between 1 and $Q$, or gets the value of a randomly selected nearest
neighbour on the lattice. These mutations, in contrast to those of the Penna
ageing model, are reversible. However, the model shows a strong first-order
phase transition with hysteresis, between dominance of one language and 
fragmentation of the population into numerous languages.  Due to this 
hysteresis the results also in this language model may depend on the 
initial conditions or more generally on history.

\begin{figure*}[!hbt]
\begin{center}
(a) \includegraphics[angle=-90,scale=0.33]{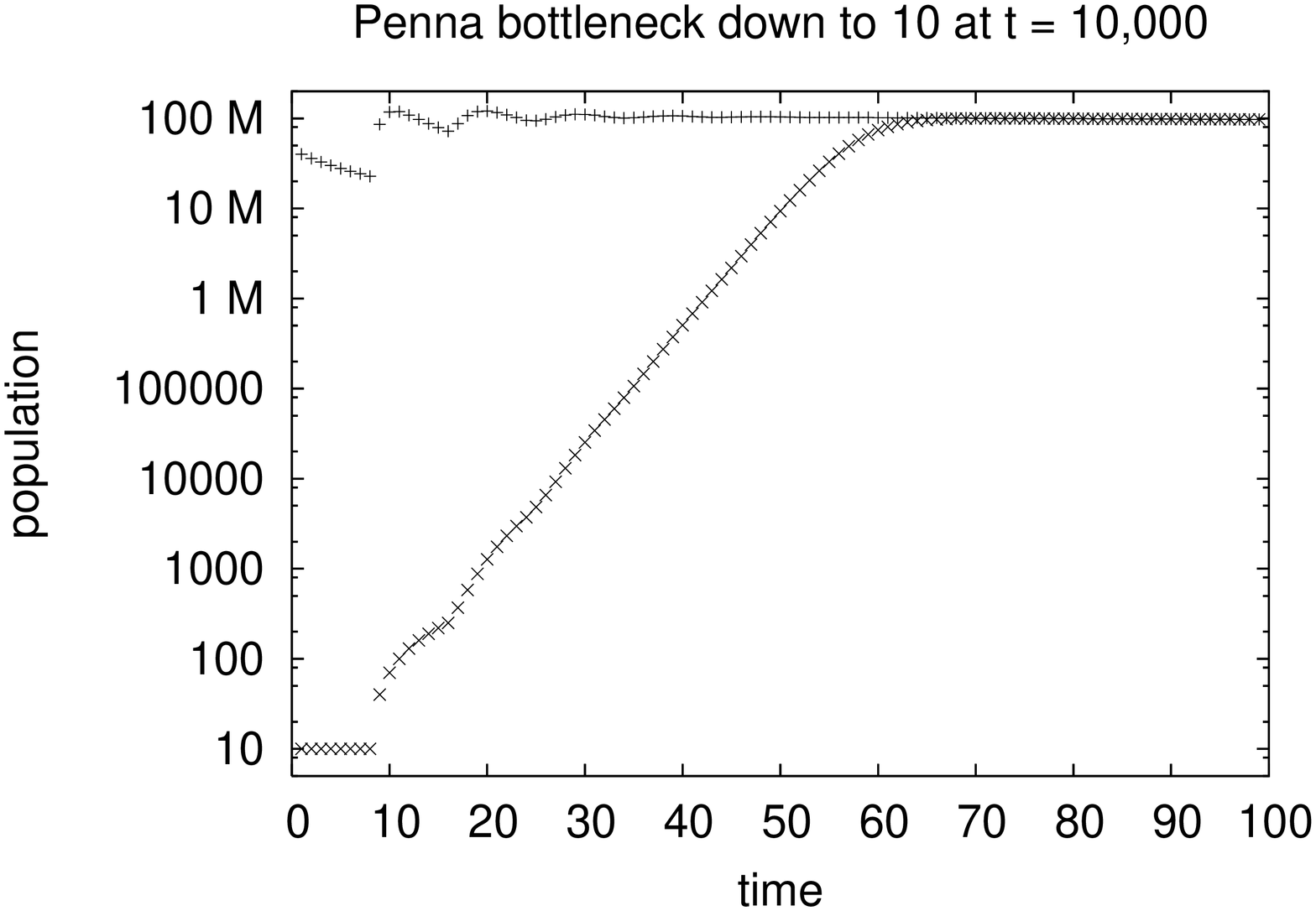}
(b) \includegraphics[angle=-90,scale=0.33]{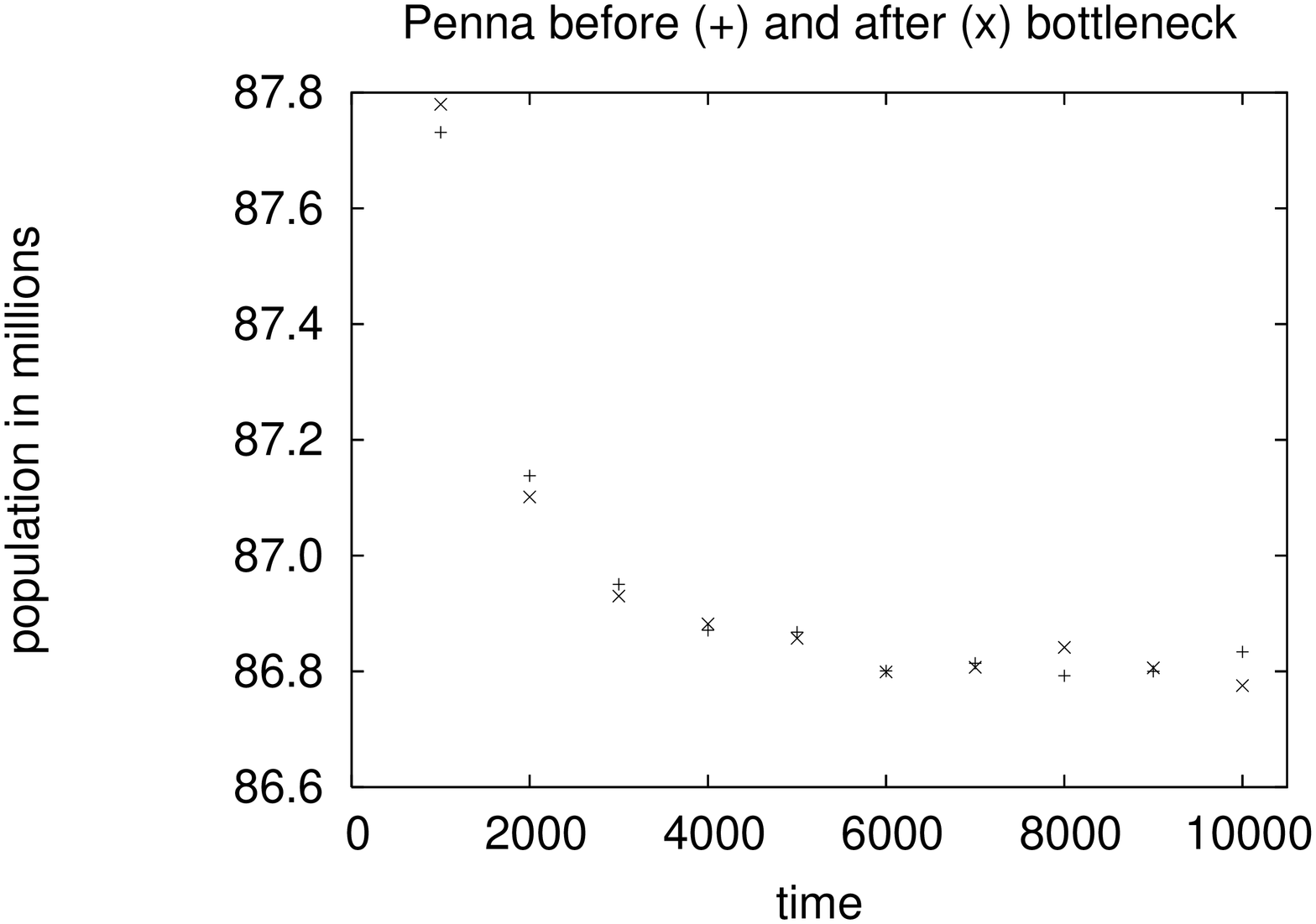}
\end{center}
\caption{Penna ageing model: Total population for short (part a) and
long (part b) times, with final agreement before (+) and after (x) the 
bottleneck.}
\label{fig-1}
\end{figure*}

\begin{figure}[!hbt]
\begin{center}
\includegraphics[width=0.45\textwidth]{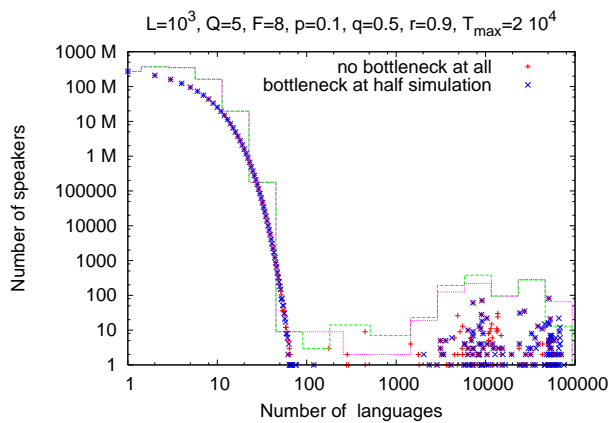}
\end{center}
\caption{Schulze language model: Histogram of language sizes on $1000 \times
1000$ lattice at mutation probability $p = 0.1$. The lines give data binned by 
powers of two in the language size = number of people speaking one language.}
\label{fig-2}
\end{figure}

\section{Simulation}
\label{sec-simulation}

In both models we made 10,000 iterations (sweeps through the population), 
then simulated a bottleneck by reducing the population to ten 
randomly selected individuals, and then made another 10,000 iterations to
find the new equilibrium. Then we compare the results after the second half
with those after the first half (before the bottleneck).

Fig. \ref{fig-1} shows for the Penna ageing model that the effects of the bottleneck
vanish after several dozen iterations, before even the total population 
gets its equilibrium value. Also in the age distribution of the population
and the distribution of 1 bits among the 32 bit positions no significant
difference was found (not shown.) 

Fig. \ref{fig-2} shows the size distribution of languages, i.e. we count how many
languages exist which are spoken by a given number of people. Again, no 
effect of the bottleneck is seen. The same holds for the time development 
of the size of the largest language (not shown). (The bottleneck consisted
in randomly selecting ten survivors and to let them
randomly populate the whole lattice again.)  
 
\section{Summary}
\label{sec-summary}

These simulations showed no long-time bottleneck effects. This does not 
exclude that with other parameters or in other models such effects 
exist. And for short times the bottleneck effects are clearly visible,
Fig. \ref{fig-1}(a). Thus, if for example the species {\it Homo Sapiens} survived a 
bottleneck $10^5$ years ago, it could be that some genes mutate with a 
probability of less than $1/10^5$ per year and thus we are not yet in
genetic equilibrium.

\begin{acknowledgments} 
We thank Y. Louzoun for suggesting this work during the GIACS Sociophysics 
summer school, September 2006, and S. Moss de Oliveira for comments on the 
manuscript.
Part of calculation was carried out in ACK-CYFRONET-AGH.
Time on HP Integrity Superdome is financed with grant MNiI/HP\_I\_SD/AGH/047/2004.
\end{acknowledgments} 

\end{document}